\documentclass[mathleft
]{an}
\usepackage{graphicx}
\usepackage{times}
\begin{document}

\Pagespan{1}{}
\Yearpublication{2010}%
\Yearsubmission{2009}%
\Month{11}%
\Volume{999}%
\Issue{88}%

\title{On the relation between supersoft X-ray sources and VY Scl stars:\\
       The cases of V504 Cen and VY Scl\thanks{Based partially on 
        CNTAC programs 04B-0086, 05A-0004, 05B-0209, 06A-0099, 06B-0250,
        07A-0384, 07B-0054, 08A-0026, 08B-0056, 09A-0025, 
        as well as ESO DDT 276.D-5040 and 277.D-5034}}

\author{J. Greiner\inst{1}\fnmsep\thanks{Corresponding author:
  \email{jcg@mpe.mpg.de}}
\and  R. Schwarz\inst{2}
\and C. Tappert\inst{3}
\and R.E. Mennickent\inst{4}
\and K. Reinsch\inst{5}
\and G. Sala\inst{6}
}
\titlerunning{SSS and YV Scl stars}
\authorrunning{J. Greiner et al.}
\institute{Max-Planck Institut f\"ur extraterrestrische Physik, 
  Giessenbachstr. 1, 85740 Garching, Germany
\and
Astrophysikalisches Institut Potsdam, An der Sternwarte 16, 
14482 Potsdam, Germany
\and 
Depart. de Fisica y Astronomia, Facultad de Ciencias, Universidad de Valparaiso, 
  Av. Gran Breta\~na 1111, Chile
\and
Depart. de Astronomia, Facultad de Ciencias Fisicas y Matematicas,
  Universidad de Concepcion, Casilla 160-C, Chile
\and
Inst. f\"ur Astrophysik, Georg-August-Univ. G\"ottingen,
  Friedrich-Hund-Platz 1, 37077 G\"ottingen, Germany
\and
Depart. Fisica i Enginyeria Nuclear, Univ. Politec. de Catalunya, 
  C. Comte d'Urgell 187, 08036 Barcelona, Spain
}

\received{26 Aug 2009}
\accepted{?? Oct 2009}
\publonline{later}


\abstract{%
We summarize our optical monitoring program of VY Scl stars with the 
SMARTS telescopes, and triggered X-ray as well as optical observations
after/during state transitions of V504 Cen and VY Scl.
}

\maketitle

\section{Introduction}

VY Scl stars are a subclass of cataclysmic variables (CVs) 
that are optically bright most of the time,
but occasionally drop in brightness by several magnitudes  at irregular
intervals (Bond 1980, Warner 1995, Honeycutt \& Kafka 2004). The transitions 
between the brightness levels take place 
\linebreak within days to weeks.
In their high states, these variables have the largest time-averaged mass
transfer rate $\dot{M}$ (of the order of 10$^{-8}$ M$_{\odot}$/yr) among CVs,
and thus are thought to be steady accretors with hot disks.
The cause of the transitions is widely debated. The conventional view is that
a strong reduction (or even cessation) of the mass transfer
rate, either due to the magnetic spot covering temporarily the $L_1$ region
(Livio \& Pringle 1994) or due to non-equili\-brium effects in the
irradiated atmosphere of the donor (Wu et al. 1995) causes the deep low states.
A major problem of this scenario is the lack of dwarf nova (DN) outbursts
during the low states (Leach et al. 1999): the disk remains subject to the 
thermal/viscous instability as it must drain its remaining gas onto the 
white dwarf (WD) through a series of DN eruptions. Another serious difficulty 
is that the
observed dual-slope rises (Honeycutt \& Kafka 2004), which are faster when the
system is fainter, are
opposite to the expected behaviour of an accretion disk since rebuilding the
disk from an extended low state should initially take place on the slow viscous
timescale of the low-state disk, eventually switching to a faster rise as the 
disk goes into DN outburst on the faster thermal timescale. 

Two unconventional views invoke quasi-stable burning of the accreted hydrogen
on the white dwarf surface or a magnetic nature of the white dwarf. In the
former case, VY Scl stars  may be transient supersoft X-ray sources during
optical low-states (Greiner \& DiStefano 1999, Grei\-ner 
\linebreak 
2000, Honeycutt 2001).
Discovering more nearby supersoft X-ray sources is potentially important,
since some of them may be progenitors of Type Ia supernovae; in addition, 
the sources 
may play an important role as ionizers of the interstellar medium. This picture
 is motivated by the behaviour of the classical supersoft X-ray source
RX J0513.9--6951 which has quasi-periodic optical intensity dips of $\sim$4
weeks duration (Southwell et al. 1996, Reinsch et al. 2000) 
simultaneously to X-ray high states.
This is interpreted as being due to a drop
in accretion rate which leads to a contraction of the WD, and in turn a hotter
WD surface temperature. Besides this phenomenological similarity, there
exist independent observational hints for transient supersoft X-ray
emission in 5 VY Scl stars:  
V751 Cyg (Greiner et al. 1999),
V~Sge (Greiner \& Teeseling 1998),
DW UMa (Knigge et al.\ 2000), 
SW Sex (Groot et al.\ 2001),
BZ Cam (Greiner et al. 2001).

This conjecture that VY Scl stars are the low-mass extension of supersoft X-ray 
binaries (SSB) can only be tested by combined optical and X-ray observations. 
If it can be 
pro\-ven, it is expected to have far-reaching consequences.
For example, if a relation to SSBs
can be established, it would add a whole group of optically bright
objects that are much easier to study than the optically-faint SSBs.
Second, because the presence of WDs in SSBs has never been proven,
a connection of SSBs and VY Scl stars would add support to the standard
model of supersoft X-ray sources (van den Heuvel et al.\ 1992).
Third, the irradiation of the donors in supersoft X-ray sources is much
stronger than in CVs (and VY Scl stars), and therefore
the mechanism proposed by Wu et al.\ (1995) for VY Scl stars could be
readily applicable to SSBs.

In the case of a magnetic primary, a low magnetic field of 
order 5$\times$10$^{30}$ G cm$^3$
would disrupt the inner, otherwise unstable accretion disk, and thus
prevent outbursts in low states, suggesting
that VY Scl stars may all be intermediate polars (Hameury \& Lasota 2002).
While variable circular polarization (and consequently a magnetic field)
has been found in two VY Scl stars (LS Peg and V795 Her), the majority has
no measurable magnetic field.
This conjecture can be "easily" tested by X-ray observations (searching
for X-ray pulsations) or optical spectropolarimetry.

Here, 10 yrs after the first suggestion (Greiner \& DiStefano 1999),
we report on our results of long-term optical monitoring of southern 
VY Scl stars, and triggered follow-up optical and X-ray observations
when a VY Scl star went into an optical low-state.

\section{The monitoring program}

We primarily selected VY Scl stars with known distance,
so that we could determine the X-ray and optical luminosity: 
PX And, TT Ari, KR Aur, MT Pup, SW Sex, ST Cha, V504 Cen,
LX Ser, V442 Oph, V794 Aql, 
LQ Peg, VY Scl and VZ Scl.
Monitoring observations were performed  2004--2009 with ANDICAM 
on the SMARTS 1.3\,m telescope in $B$,$R$ and $J$ once every $\approx$4 nights 
throughout their visibility periods. 
For the occurrence of optical state transitions, we were prepared by
either pre-approved or ready-to-submit target-of-opportunity proposals
to obtain X-ray observations, optical spectroscopy as well as
optical spectro-polarimetry. 

Two objects of our sample exhibited an optical state transition, and these
are described below in more detail.

\begin{figure}[th]
\hspace{-0.5cm}
\includegraphics[width=0.79\columnwidth,angle=270]{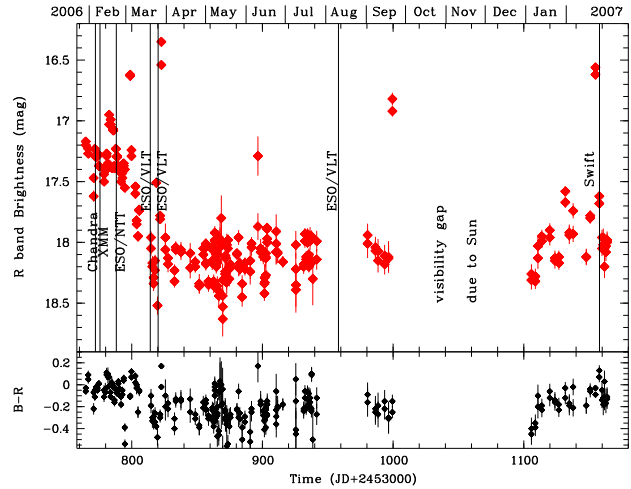} 
\caption{Light curve of V504 Cen in the R band (top) and its $B-R$ 
color (bottom) as obtained with our SMARTS program
since January 2006. The typical
high-state brightness is 12.2 mag. The scatter in the low-state is most
likely completely due to the orbital variability (see Fig. 2). 
Overplotted are the
times at which Directors Discretionary Time (DDT) or Target-of-Opportunity
(ToO) observations at other facilities were triggered. Note the color
change when going from the intermediate state of 17.3 to the low-state
at 18.3 mag in Mar. 2006. The Swift ToO was attempted quickly after the 
recognition of the brightening, which turned out to be a flare and not 
  the rise back to the bright state.
V504 Cen remained in its low-state since then, and still is
as of April 2009, the end of our monitoring campaign. 
This is particularly long for VY Scl stars.\label{olc}
}
\end{figure}

\begin{figure}
\hspace{-0.3cm}
\includegraphics[width=0.99\columnwidth]{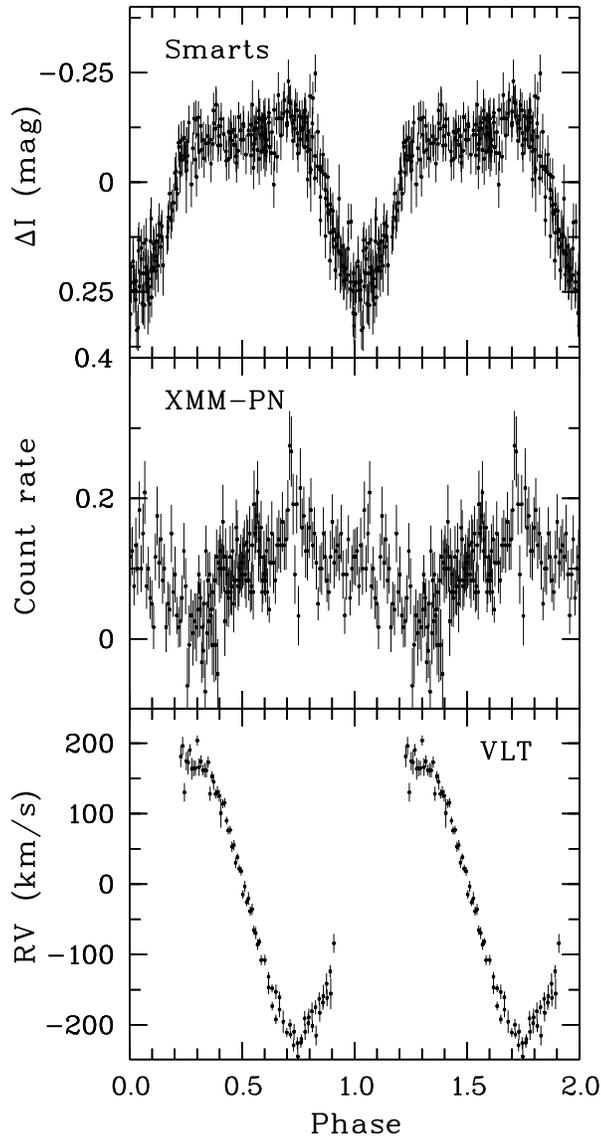}
\caption{{\bf Top:} All SMARTS I-band photometric data of V504 Cen folded 
over our best fit period of 252.816$\pm$0.020 min $\equiv$4.21 hrs.
{\bf Center:} 
XMM-pn light curve folded over the same period.
Note the distinct shift of the X-ray versus the optical minimum.
{\bf Bottom:} Radial velocity curve of the iron lines (Fig. \ref{ospec}).
The optical minimum coincides with the blue-red zero-crossing.
However, the width of nearly 0.4 phase 
units is too large for the occultation of an accretion disk by the companion.
\label{foldlc}
}
\end{figure}

\subsection{V504 Cen}

V504 Cen (also AN 354.1935) is a poorly studied CV, and was proposed
as a member of the VY Scl class by Kilkenny \& Lloyd Evans (1989). 
The first-ever well-covered fading episode
was reported by Kato \& Stubbings (2003) based on
data of the all-sky survey ASAS-3 of 2002. The duration of the
fading episode was about 200 days, with an amplitude of $>$2 mag
(only upper limits were obtained).
No distance or orbital period has been reported for V504 Cen.

V504 Cen has been monitored at the 1.3m SMARTS telescope in CTIO/Chile
since 2004. Folding the long-term monitoring data from the low-state
(excluding the Jan-Mar data because of the higher intensity level),
we obtain an orbital period of 4.21 hrs (Fig. \ref{foldlc}), with the
following epheme\-ris: T$_{\rm min}$ = 0.17556655$\times$ E + 2453909.61279. 
This period is confirmed (Fig. \ref{foldlc}) 
with 2--3 hrs phase-resolved I-band imaging in each of 5 nights,
spread through 2006 and 2007.

When VSNET 
on 26 Jan. 2006 reported a drop in brightness, our monitoring 
had just resumed after the visi\-bi\-lity-break due to the nearby Sun,
and observations (in 3 colors) confirmed the fading: the 
source had gone from B = 12 mag (last observation 19. Sep. 2005) 
down to B$\sim$17.0 mag, and 
later went even further to  B$\sim$18.0 mag in March 2006.
We subsequently triggered 
observations with Chandra, XMM-{\it Newton}, VLT and NTT (ESO), and Swift 
(Fig. \ref{olc}).

The Chandra observation on Feb. 6, 2006 clearly detected 
V504 Cen at 0.068$\pm$0.005 cts/s, however not with a soft X-ray spectrum.
Photons up to 2 keV are detected, and the data are best fit with a
power law model with photon index 3.2$\pm$0.5, and foreground absorption
consistent with zero ($\chi_{red}$ = 1.1). Bremsstrahlung (kT = 0.4 keV)
or blackbody (kT = 0.14 keV) models fit substantially worse.
We did not detect a soft X-ray spectrum 11 days after the first
report of optical faintness; this time span is a lower limit since the 
beginning of the optical low-state is unobserved.

The XMM-{\it Newton} observation was performed on Feb. 9, 2006 for 
20 ksec. Photons are detected up to 9 keV, and the spectrum is best
fit with a two-temperature MEKAL model, with 0.83 keV and 5.64 keV.
Also the 6.7 keV iron line is clearly detected.
The best-fit absorbing column is N$_{\rm H}$ = 6.5$\times$10$^{20}$ cm$^{-2}$,
corresponding to the foreground galactic value (Dickey \& Lockman 1990).
The unabsorbed 0.2--10.0 keV flux is 
 4.3$\times$ 10$^{-13}$ erg cm$^{-2}$ s$^{-1}$, corresponding to
1.2$\times$ 10$^{31}$ (D/500 pc)$^2$ erg/s.
No X-ray pulsations with period longer than 4 min and with amplitude
larger than $\sim$5\% were detected.

\begin{figure}
\hspace{-0.1cm}
\includegraphics[width=0.40\columnwidth,angle=270]{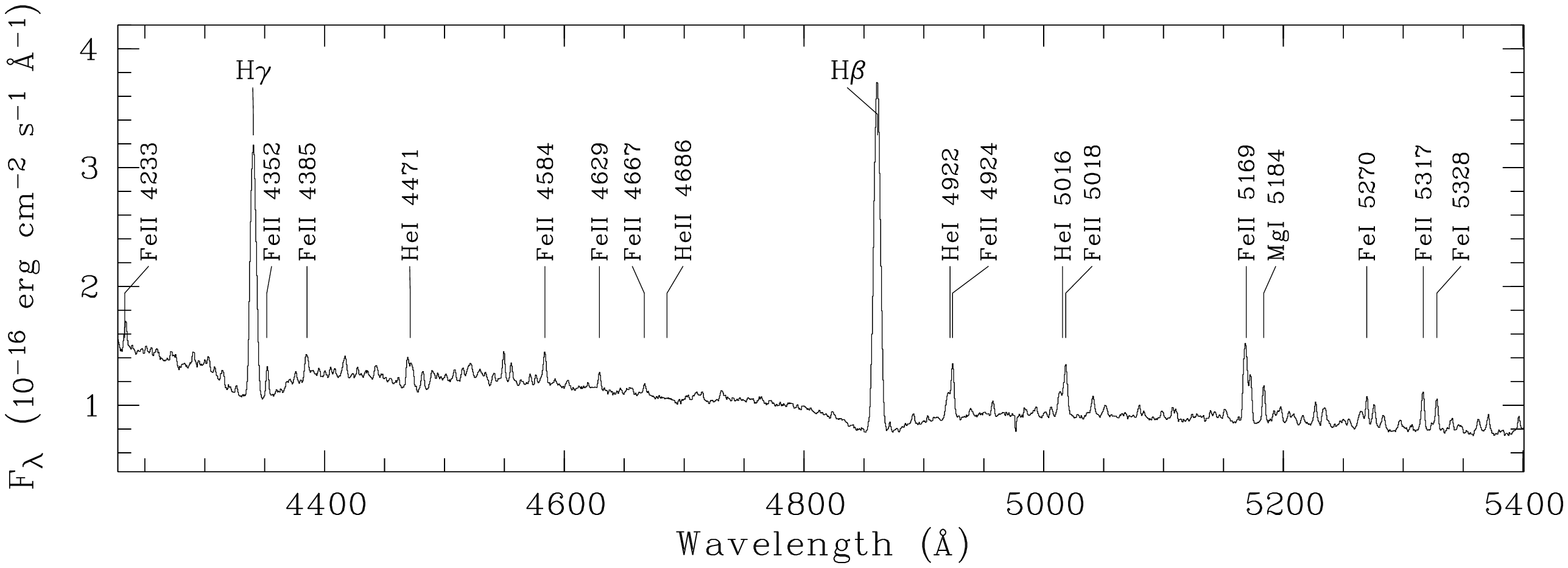}
\includegraphics[width=0.73\columnwidth,angle=270]{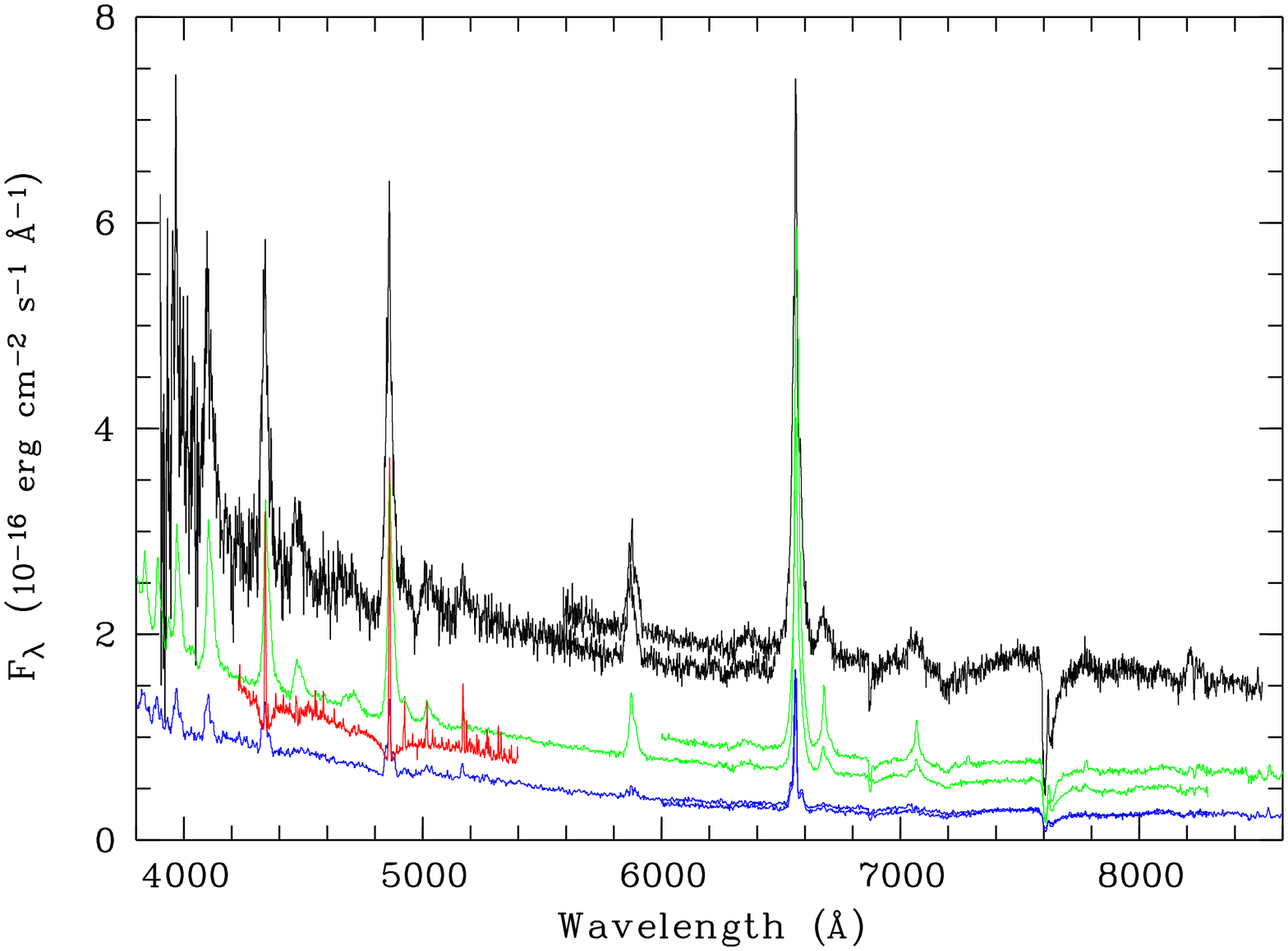}
\caption{Spectra of V504 Cen (bottom panel) obtained with EMMI/NTT on Feb. 10, 
2006 (black)
and FORS1/VLT with different grisms on March 10 and 21-24, 2006.
The mean spectrum (top panel) shows a wealth of low-ionization Fe lines,
most likely from the heated side of the donor. No HeII 4686 \AA\ is detected.
\label{ospec}
}
\end{figure}

Spectra were obtained with the ESO NTT on Feb 10, 2006 with grating \#8
to cover the red part of the spectrum. The mass donating star is clearly
identified; the spectral type is M3$\pm$1. The donor contributes
about 20\% of the light in the red during this time (m$_{\rm R} \sim$19 mag).
With the absolute 
flux calibration and using the semi-empirical absolute magnitude of
CV secondaries (Knigge 2006), in our case M$_{\rm R}$ = 9.57 mag, this
places V504 Cen at a distance of 600 pc.

The ESO/VLT observations were performed at two occasions: Firstly,
5 hrs phase-resolved FORS1 spectroscopy was done March 20--24, 2006.
Secondly, FORS1 was used in spectro-polarimetry mode on Aug. 11, 2006.
The March observations reveal a wealth of FeII lines, strong Balmer lines
in emission, but no HeII (Fig. \ref{ospec}). 
Both the iron as well as the Balmer lines
were used to obtain a spectroscopic period of 4.21 hrs
(coincident with the photometric period). 
The spectro-polarimetry with FORS1/600B was compromised by
mediocre conditions. 
We did not detect circular polarization in the mean spectrum
at a level higher than 3\% (3$\sigma$ level).
This level of polarization is substantially above
the range of polarization we were aiming for to detect, namely 0.1\%. 
Unfortunately, 
therefore, the result is inconclusive, since the other two 
VY Scl stars with reported circular polarization have 0.3\%.

\begin{figure}
\hspace{-0.5cm}
\includegraphics[width=0.73\columnwidth,angle=270]{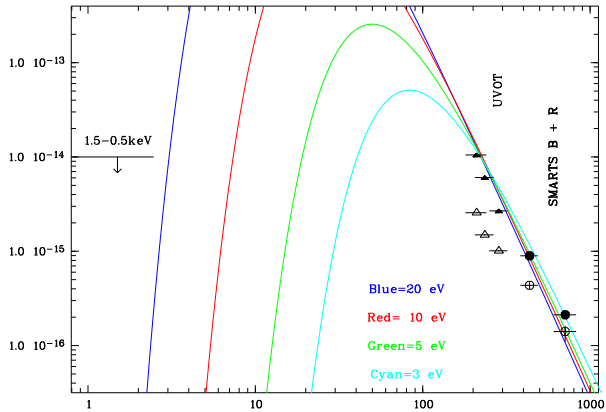}
\vspace{-0.2cm}
\caption{Spectral energy distribution of V504 Cen 
from quasi-simultaneous observations of Swift and SMARTS
on Feb 26, 2007. No X-rays are detected (upper limit shown as arrow), 
and the Swift/UV
and SMARTS data are consistent with a 35\,000 K (3 eV) accretion disk, hotter
than the canonical 10\,000 K. 
\label{sed}
}
\end{figure}

The Swift observation was performed on Feb. 26, 2007, between 20:31--22:37 UT,
in response to a 1.5 mag optical brightening the day earlier. We attempted
quasi-simulta\-neous SMARTS observations, which in the end happened 79 min 
before the X-ray observation.
V504 Cen is not detected at X-rays, but clearly in the UV filters.
Applying an extinction correction of A$_V$=0.59 mag, the UVOT and SMARTS
data nicely follow the Wien-tail of a blackbody spectrum. The X-ray
upper limit constrains the temperature of any hot component in the
system to kT$<$20 eV (Fig. \ref{sed}).

\subsection{VY Scl}

\begin{figure}
\hspace{-0.5cm}
\includegraphics[width=0.73\columnwidth,angle=270]{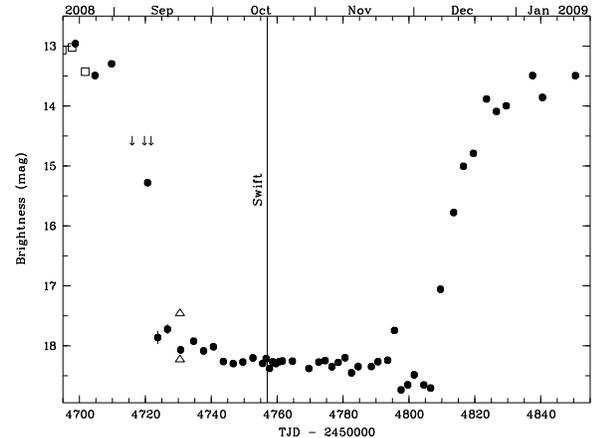}
\caption{B band light curve of VY Scl (filled symbols) as obtained 
with our SMARTS program.
Open squares and upper limits (arrows) are from ASAS (reported via VSNET),
while the two open triangles are $V$-band CCD observations
(J. Temprano \& Cantabria Obs). 
\label{vyscl}
}
\end{figure}

The prototype of the class, VY Scl, started to transition into
an optical low state in early September 2008. Our SMARTS coverage
is reasonable, and shows that the transition happened within less than 
12 days (Fig. \ref{vyscl}). There is a further drop 
in intensity after another 15 days.  
The $\sim$10 day
period of even lower intensity just before the (slow) rise into the high state
is also remarkable.
 This was the first deep low-state of VY Scl since 1983 
(Cropper and Warner  1983), after a shallow one ($B \sim$15.5 mag)
reported via VSNET in Dec. 2003. The low-state ended after only
3 months.

A 10 ksec Swift ToO observation was obtained a week after reaching the 
B $\approx$ 18.2 mag base-level. Only 37 photons are detected, which are
spread in energy up to 6 keV.
The observed 0.5--10 keV X-ray flux is 
1.2$\times$ 10$^{-12}$ erg cm$^{-2}$ s$^{-1}$, corresponding to a luminosity of
3.5$\times$ 10$^{31}$ (D/500 pc)$^2$ erg/s. This is slightly smaller than 
the upper limit
obtained from ROSAT observations in 1990 (Greiner 1998).

\section{Summary}


For V504 Cen, we tested for a hot source in March 2006. 
The optical spectra do not show any significant
sign of HeII emission, and Chandra and XMM ToO observations
do not show a strong supersoft X-ray component.
The magnetic option was tested with spectro-polarimetry, but no sign
of a substantial magnetic field was found.
%
%
It thus appears that the suspected H surface burning during the low-states
in VY Scl stars either happens at very low temperatures (below 20 eV),
or is not a generic feature in this class of objects.
Similarly, a magnetic nature of the WD could not be proven.
This indicates that in V504 Cen
both scenarios proposed as ``unconventional'' views could not be 
substantiated, 
but also the conventional Lagrange point blockage model is challenged by
the very long duration of the optical low-state.

\acknowledgements
We are grateful to many people who helped making all these correlated
observations happen: J. Nelan, M. Buxton, S. Tourtellotte (all SMARTS),
H. Tananbaum, A. Prestwich, N. Adams-Wolk (all Chandra),
N. Schartel, N. Loiseau (both \linebreak XMM),
N. Gehrels, M. Chester (both Swift), 
M. Arnaboldi, E. Pompei, V. Ivanov, M. Rejkuba (all ESO).
We also thank N. Prymak for data reduction of the SMARTS data of V504 Cen 
during the first year of the project.




\end{document}